\documentclass[prl,groupedaddress,twocolumn,showpacs]{revtex4}
\usepackage[dvips]{graphicx}

\begin{document}

\title{Evidence for pseudogap and phase-coherence gap separation by Andreev
reflection experiments in Au/La$_{2-x}$Sr$_x$CuO$_4$ point-contact
junctions}
\author{R.\@S. Gonnelli}
\email[Corresponding author. E-mail:]{gonnelli@polito.it}
\affiliation{INFM - Dipartimento di Fisica, Politecnico di Torino,
10129 Torino, Italy}
\author{A. Calzolari}
\affiliation{INFM - Dipartimento di Fisica, Politecnico di
Torino, 10129 Torino, Italy}
\author{D. Daghero}
\affiliation{INFM - Dipartimento di Fisica, Politecnico di
Torino, 10129 Torino, Italy}
\author{L. Natale}
\affiliation{INFM - Dipartimento di Fisica, Politecnico di
Torino, 10129 Torino, Italy}
\author{G.\@A. Ummarino}
\affiliation{INFM - Dipartimento di Fisica, Politecnico di
Torino, 10129 Torino, Italy}
\author{V.\@A. Stepanov}
\affiliation{P.N. Lebedev Physical Institute, Russian Academy of
Sciences, 117924 Moscow, Russia}
\author{M. Ferretti}
\affiliation{Dipartimento di Chimica e Chimica Industriale,
Universit$\grave{a}$ di Genova, 16146 Genova, Italy}

\date{\today}

\begin{abstract}
We present new Au/La$_{2-x}$Sr$_x$CuO$_4$ (LSCO) point-contact
conductance measures as a function of voltage and temperature in
samples with $0.08 \leq x \leq 0.2$. Andreev reflection features
disappear at about the bulk $T_{\mathrm{c}}$, giving no evidence
of gap for $T > T_{\mathrm{c}}$. The fit of the normalized
conductance at any $T < T_{\mathrm{c}}$ supports a ($s+d$)-wave
symmetry of the gap, whose dominant low-$T$ \emph{s} component
follows the $T_{\mathrm{c}}(x)$ curve in contrast with recent
angle-resolved photoemission spectroscopy and quasiparticle
tunneling data. These results prove the separation between
pseudogap and phase-coherence superconducting gap in LSCO at $x
\lesssim 0.2$.
\end{abstract}
 \maketitle

In a recent paper, G. Deutscher claimed the existence of two
distinct energy scales - that is, two distinct gaps~- in
high-$T_{\mathrm{c}}$ superconductors (HTS) \cite{ref1}. According
to his discussion, one of these gaps should appear at $T^{*} >
T_{\mathrm{c}}$ in optimally-doped and underdoped samples and
could be due to an incoherent pairing between charge carriers
(whose physical origin is still under discussion) which leads to a
pair pre-formation. This gap, $\Delta_{\mathrm{p}}$, would
coincide with the pseudogap observed by angle-resolved
photoemission spectroscopy (ARPES) and tunneling experiments. The
second gap, $\Delta_{\mathrm{c}}$, would appear at
$T_{\mathrm{c}}$ and would be associated to the achievement of the
phase coherence by the pre-formed pairs and, consequently, to the
onset of superconductivity. This \emph{phase-coherence} gap can be
observed only by experimental tools sensitive to the phase
coherence of the pairs, i.e. Josephson effect and/or Andreev
reflection experiments. At the present moment low-temperature
tunneling \cite{ref4} and very recent ARPES experiments
\cite{ref6} in La$_{2-x}$Sr$_x$CuO$_4$ (LSCO) have shown at
$x<0.2$ the presence of a large gap which increases at the
lowering of the doping level. Very few experiments have instead
been performed to investigate the Andreev gap \cite{ref6b,ref8},
and, to our knowledge, none at all to study in detail its
dependence on the temperature and on the doping in the region from
overdoped to underdoped.

In this letter, we present and discuss the results of
point-contact experiments on La$_{2-x}$Sr$_x$CuO$_4$ samples.
Despite the polycrystalline nature of the samples, a very careful
point-contact technique allowed obtaining reproducible Andreev
reflection curves and studying for the first time their behaviour
in a broad temperature and doping range. In order to extract
information about the dependence of the Andreev gap on $x$ and
$T$, we fitted the experimental curves with the generalized BTK
model by Y.~Tanaka and S.~Kashiwaya \cite{ref12} for various
possible symmetries of the order parameter. We found that the
dependence of the Andreev gap on temperature and Sr content
experimentally proves the existence of two distinct energy scales,
a large pseudogap and a smaller superconducting gap, in LSCO.

The high-quality La$_{2-x}$Sr$_x$CuO$_4$ polycrystalline samples
used in our measurements were prepared by conventional solid-state
reaction at 1000$^{\circ}$C by using stoichiometric amounts of
the high-purity precursor oxides La$_2$O$_3$, CuO, and SrO$_2$.
After the first reaction step the bulk materials were finely
ground, pressed into small rectangular bars and sintered to
obtain higher density samples. The sintering temperature was
selected between 1100° and 1150$^{\circ}$C for different Sr
amount. When the dopant concentration was greater than $x$=0.1,
quenching was required from higher temperatures (1170$^{\circ}$C)
to ensure chemical homogeneity. All samples were structurally
characterized by XRD powder diffraction \cite{ref9}, and their
actual stoichiometry was determined by means of EDS microprobe
analysis, which evidenced the absence of impurities and confirmed
their nominal Sr concentrations: $x=$ 0.08, 0.10, 0.12, 0.13,
0.15 and 0.20. The typical linear dimension of the grains, as
observed by means of AFM or SEM measurements, was 5$\div 10\;
\mu$m. AC susceptibility and resistivity measurements were used
to determine the critical temperatures, which were in good
agreement with the standard curve of $T_{\mathrm{c}}$ as a
function of $x$ for LSCO \cite{ref10}. The width of the resistive
transition was of the order of 3$\div$5~K for all the Sr contents.

We performed on these samples point-contact experiments with Au
tips, whose ending-part diameter was always less than $\sim 2 \;
\mu$m \cite{ref10b}, obtained by electro-chemical etching (with a
HNO$_3$+HCl solution) of a 0.2 mm diameter Au wire. We often
obtained SN junctions with clear Andreev reflection
characteristics. Due to the stability of the point contacts, we
were able to follow the evolution of the conductance curves on
heating the junction from 4.2~K up to the temperature
$T_{\mathrm{c}}^{\mathrm{A}}$ at which the dynamic conductance
d$I$/d$V$ was flat.

\begin{figure}[t]
\vspace{-3mm}
\begin{center}
\includegraphics[keepaspectratio,width=0.7\columnwidth]{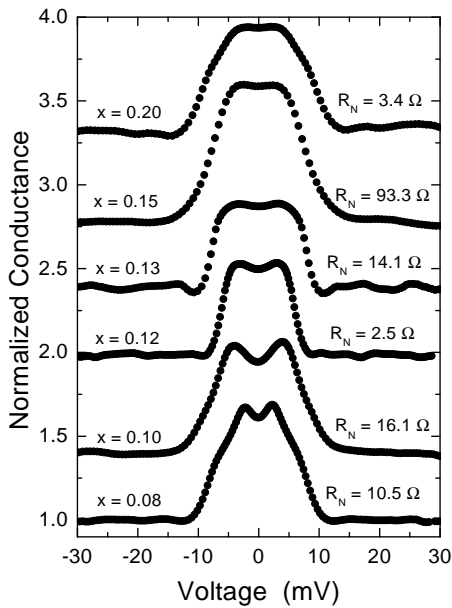}
\vspace{-6mm}\caption{\small{The normalized conductances of
Au/LSCO point-contact junctions for various doping levels ($0.08
\leq x \leq 0.2$) and at low temperature (4.22 K $\leq T \leq$
5.61 K). The curves are vertically displaced for
clarity.}}\vspace{-5mm}
\end{center}
\end{figure}

Figure~1 shows the low-temperature experimental normalized
conductance data (vertically shifted for clarity) for the six
doping values previously mentioned. We systematically normalized
only the data sets for which d$I$/d$V$ at $|V| >$ 20 mV was
reasonably constant and did not show sensible variations at the
change of temperature. All the results that we show in the present
letter are obtained from this kind of data.

The normal-state resistances of the junctions for all dopings are
indicated near the curves in Fig.~1. With these contact
resistances, and with the estimation of $k_{\mathrm{F}}$ (from
$E_{\mathrm{F}} \sim 100$~meV) and of the mean free path (from
$k_{\mathrm{F}}\ell \approx 13$ \emph{at the transition
temperature} as reported in Ref.\cite{Boebinger}) one obtains that
the contact radius $a$ ranges from 146 $\mathrm{\AA}$ (when $R%
\sim 90\, \Omega$) to about 800 $\mathrm{\AA}$ (when $R \sim 3 \,
\Omega$), while $\ell$ ranges from 40 to 70 $\mathrm{\AA}$ from
underdoped to overdoped. Then, if single contacts are established
between the tip and the material under study, they are \emph{not}
in the Sharvin limit. On the other hand, the $I-V$
characteristics give no evidence of heating phenomena. In fact,
the variation of conductance with bias is within that expected in
the ballistic regime \cite{Srikanth} and much smaller than that
expected if the junction was heated up to a bias-dependent
temperature above the bath one. Thus, we can exclude to be in the
Maxwell (thermal) regime, which is enough to ensure that the
conditions for energy-resolved spectroscopy are fulfilled, as
widely shown in literature \cite{Jansen}. The logical
consequence, also supported by the polycrystalline nature of our
samples and by the softness of the Au tip, is that the low
contact resistances can be explained by the presence of several
parallel ballistic contacts between sample and tip \cite{Aminov}.

Thus, the features we observed in the experimental data of Fig.~1
are with no doubt due to Andreev reflection at the S-N interface.
Nevertheless, some differences are present with respect to the
ideal curves predicted for a very low potential barrier by the
well-known BTK model \cite{ref10c}. The maximum value is less than
that expected and the shape is not always compatible with a pure
\emph{s}-wave symmetry of the order parameter. Moreover, some more
or less pronounced oscillations of d$I$/d$V$ are present at $|V|
\gtrsim $ 10 mV. These oscillations have been already observed in
HTS and can be due to the presence of localized electron states in
the interface potential barrier \cite{ref11b}.

\begin{figure}[t]
\vspace{-10mm} \hspace{-3mm}
\includegraphics[keepaspectratio,width=\columnwidth]{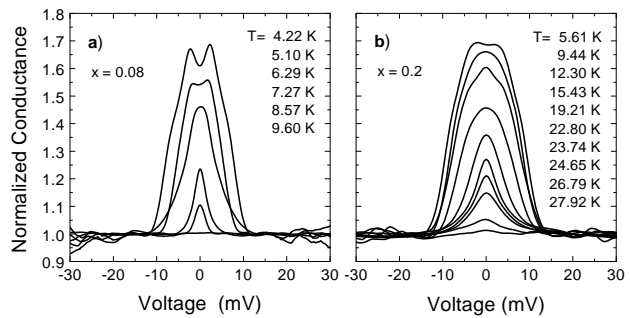}
\vspace{-10mm} \caption{\small{Temperature dependence of the
normalized Andreev conductance in LSCO samples with $x=$ 0.08 (a)
and 0.2 (b).}} \vspace{-2mm}
\end{figure}

Figure~2 shows the temperature dependence of the normalized
conductance in samples with $x$\@=\@0.08 (a) and $x$\@=\@0.2 (b).
As in all the other samples, the Andreev features disappear at
the bulk $T_\mathrm{c}$ (or at a slightly lower temperature) and
a change in the shape of the curves is evident at the increase of
$T$ (see, for example, the curve at $T$\@=\@12.3~K in Fig. 2b).
As we will show later, this last feature can be explained by a
change in the relative weight of the isotropic and anisotropic
gap components.

To evaluate the gap and to study its dependence on the doping
content, we fitted the normalized conductance curves by using the
generalized BTK model introduced some years ago by S.~Kashiwaya
and Y.~Tanaka \cite{ref12}. In order to properly fit our data in
the whole temperature range, we introduced in the original model
of Ref.~\cite{ref12} the effect of the temperature and of the
broadening parameter $\Gamma$ which takes into account the finite
lifetime of the quasiparticles. Various symmetries of the order
parameter were used (\emph{s}, \emph{s}+i\emph{d},
\emph{s}+\emph{d} and anisotropic \emph{s}). The pure
$d_\mathrm{{x^2-y^2}}$ symmetry was not considered because it was
unable to properly fit the low-voltage part of all our
low-temperature data for any value of the fit parameters.

In the case of mixed pair symmetry and at constant $T$ the free
parameters of the fit are: the values of the isotropic and
anisotropic components of the gap ($\Delta_{\mathrm{is}}$ and
$\Delta_{\mathrm{an}}$), the parameter $Z$ (proportional to the
potential barrier height), the lifetime broadening $\Gamma$ and
the angle $\alpha$ between the $a$ axis and the normal to the S-N
interface \cite{ref12}. Actually, when $Z\leq 0.3$ (as in our
case) the choice of $\alpha$ has a negligible influence on the
values of $\Delta_{\mathrm{is}}$ and $\Delta_{\mathrm{an}}$
determined by the fit, independently of the symmetry used.
Therefore, $\alpha$ is \emph{not} a critical parameter and thus we
put $\alpha=0$ in all cases. $Z$ was determined by the fits in
the various symmetries of the lowest-temperature curves and, due
to the stability of the contact resistances, was supposed to
remain constant at the increase of the temperature. The parameters
$\Delta_{\mathrm{is}}$, $\Delta_{\mathrm{an}}$ and $\Gamma$ were
varied in order to fit the data, but always keeping $\Gamma$ as
small as possible.

\begin{figure}[t]
\vspace{-3mm}
\begin{center}
\includegraphics[keepaspectratio,width=\columnwidth]{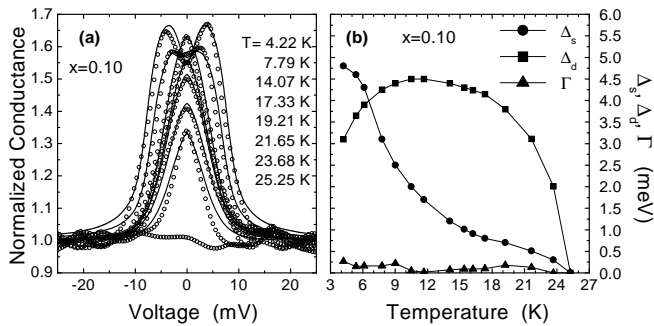}
\vspace{-17mm} \caption{\small{(a) Normalized conductance curves
at various temperatures up to $T_{\mathrm{c}}^{\mathrm{A}}$ in
LSCO with $x=$ 0.1 and best fit curves in \emph{s}+\emph{d}
symmetry. (b) Temperature dependencies of the $s$ and $d$ gap
components and of the lifetime parameter $\Gamma$ determined by
the fits of Fig. 3a. }}
\end{center} \vspace{-6mm}
\end{figure}

The theoretical conductance curves which best fit the
low-temperature experimental data of Fig.~1 are those calculated
in the (\emph{s}+\emph{d})- or in the pure \emph{s}-wave
symmetry, depending on the Sr content. In all cases, the
\emph{s}-wave component is dominant. It is very important to
notice that the value of the isotropic component of the gap is
actually \emph{almost independent} of the particular symmetry
used for the fit, and therefore can be considered a very robust
parameter.

A good fit of the conductance curves \emph{in the whole
temperature range} can be obtained only by using the
(\emph{s}+\emph{d})-wave symmetry with suitable (and
$T$-dependent) weights for the \emph{s} and \emph{d} components.
Figure 3a shows an example of the temperature dependence of the
normalized conductance in a sample with $x=$ 0.1 (open symbols)
and the corresponding (\emph{s}+\emph{d})-wave best-fit curves
(solid lines). For clarity, only few of the measured curves are
shown. The fits are good up to the critical temperature of the
junction ($T_\mathrm{c}^\mathrm{A} \approx $~25.3~K) at which the
Andreev features disappear. Similar results (always in
(\emph{s}+\emph{d})-wave symmetry) have been obtained in all the
other LSCO samples. Incidentally, Figure~3b reports the
temperature dependence of the \emph{s} and \emph{d} components of
the gap in LSCO with $x=$ 0.1 determined by the fits of Fig.~3a.
It is clear that the temperature dependence of the two components
is quite different. The shape of the $\Delta_{\mathrm{s}}(T)$
curve (and of the $\Delta_{\mathrm{d}}(T)$ one when the $d$
component is present) is common to all the doping contents.
Further details are presented elsewhere~\cite{ref13}.

Let us now go back to the discussion of the low-temperature
conductance curves shown in Fig.~1. The results of their fits are
consistent with those obtained in LSCO by Deutscher \emph{et
al.}~\cite{ref6b}. Table~I shows the temperature of the junction
and the values of $\Delta_{\mathrm{s}}$, $\Delta_{\mathrm{d}}$,
$\Gamma$ and $Z$ for the curves of Fig.~1 together with the
Andreev critical temperature $T_\mathrm{c}^\mathrm{A}$ and
$2\Delta_{\mathrm{s}}/k_{\mathrm{B}}T_\mathrm{c}^\mathrm{A}$ for
every doping value. 

\begin{table}[b]
\vspace{-4mm}
\begin{center}
\begin{tabular}{|c|c|c|c|c|c|c|c|}
\hline
  \raisebox{-1.2ex}[0pt][0pt]{Doping} & $\phantom{T}T\phantom{T}$ & $\Delta_{\mathrm{s}}$ & $\Delta_{\mathrm{d}}$ & %
  $\Gamma$ & \raisebox{-1.2ex}[0pt][0pt]{$\phantom{T}Z\phantom{T}$} & %
  \raisebox{-0.3ex}[0pt][0pt]{$\phantom{T}T_\mathrm{c}^\mathrm{A}\phantom{T}$} &
  \raisebox{-0.5ex}[0pt][0pt]{$2\Delta_{\mathrm{s}}$\raisebox{-1ex}{/}%
  \raisebox{-2ex}{$k_\mathrm{B}T_\mathrm{c}^\mathrm{A}$}}
  \\  & (K) & (meV) & (meV) & (meV) & & (K) &
  \\\hline
  0.08 & 4.22 & 3.4 & 2.5 & 0.19 & 0.20 & 9.6 & 8.2 \\ \hline
  0.10 & 4.22 & 4.8 & 3.1 & 0.27 & 0.23 & 25.3 & 4.4 \\ \hline
  0.12 & 4.22 & 5.6 & 0 & 0.92 & 0.18 & 26.0 & 5.0 \\ \hline
  0.13 & 4.22 & 6.8 & 0 & 1.50 & 0.17 & 29.1 & 5.4 \\ \hline
  0.15 & 4.65 & 6.8 & 0 & 0.44 & 0.08 & 35.3 & 4.5 \\ \hline
  0.20 & 5.61 & 6.0 & 3.5 & 1.00 & 0.13 & 27.9 & 5.0 \\ \hline
\end{tabular}
\caption{\small{Best-fit parameters and temperatures for the
curves of Fig.~1}}
\end{center}
\vspace{-3mm}
\end{table}

In Fig.~4 the doping dependencies of the \emph{low-temperature}
$\Delta_{\mathrm{s}}$ and $\Delta_{\mathrm{d}}$ (solid circles and
solid squares, respectively) determined from the data of Fig.~1
are compared to those of the ARPES leading-edge shift (LE)
recently determined in LSCO \cite{ref6} (open circles) and of the
gap determined by tunneling measurements (open
squares)\cite{ref4}. Both the ARPES LE and the tunneling gap
values increase monotonically at the decrease of the doping and
reach very large values ($15 \div 20$ meV for the ARPES LE in
strongly underdoped samples). On the contrary, the dominant
isotropic gap component determined from Andreev reflection data
increases at the decrease of the doping in the overdoped region up
to a maximum approximately located at the optimum doping, and then
strongly reduces in the underdoped region, following the critical
temperature behaviour (thick solid line). Let us stress that this
conclusion \emph{does not} depend on the model used to fit the
experimental data, and holds true even if the Andreev gap is
simply identified with the energy at which the conductance at
negative (positive) bias has the maximum (minimum) slope. Also
notice that the value of $\Delta_{\mathrm{s}}$ for $x=0.2$ almost
coincides with that measured by tunneling, and this further
supports our results, even for low contact resistances.

The main findings that follow from the results shown above can be
so summarized: i) all the Andreev reflection features disappear at
about the bulk $T_\mathrm{c}$ of the samples (see Fig. 4, open
triangles). The Andreev spectroscopy thus gives no evidence of gap
at $T > T_\mathrm{c}$ in LSCO, even in the underdoped region; ii)
the fit of the Andreev curves for all $x$ values indicate that, at
\emph{low-temperature}, the \emph{s}-wave component of the gap is
dominant and independent of the symmetry used for the fit. Pure
\emph{d}-wave symmetry is unable to fit the data; iii) in contrast
with the ARPES leading-edge shift \cite{ref6} and the gap
determined by tunneling \cite{ref4}, the low-temperature dominant
Andreev $\Delta_{\mathrm{s}}$ decreases at the decrease of $x$ in
the underdoped region and globally follows the $T_\mathrm{c}$ vs
$x$ behaviour.

These results give a complete experimental evidence for the
existence of two energy scales in LSCO. The smallest one
represents the phase-coherence (superconducting) gap, while the
greatest is related to the gap-like features (pseudogap) observed
by ARPES and quasiparticle tunneling experiments. As shown in Fig.
4, these two energy scales seem to merge slightly above the
optimum doping. The present results are also a direct prove that
the pseudogap is a property of the non-superconducting state of
LSCO. The question arises of what could be its origin. Despite the
large number of theoretical models proposed, the answer is still
not clear.

Very recently, a two-gap model appeared in literature \cite{ref14}
which explains the pseudogap features in underdoped cuprate
superconductors in the framework of incoherent pre-formed pairs
around the M points of the Brillouin zone. According to this
model, a bifurcation at $x_\mathrm{b}>x_{\mathrm{opt}}$ is
expected between the mean-field $T_\mathrm{c}$ curve (which has a
maximum at $x=x_{\mathrm{opt}}$) and the temperature of pair
pre-formation $T^{*}$ (assumed to be linearly increasing at the
lowering of $x$). Another recent model \cite{ref15}, on the
contrary, analyzes the transition to the superconducting state in
the presence of a preformed normal-state pseudogap resulting from
interactions in the particle-hole channel, and predicts for the
superconducting gap and the ARPES leading-edge shift the same
doping dependence as $T_\mathrm{c}$ and $T^{*}$ respectively, in
very good agreement with our experimental results. In conclusion,
both these approaches seem able to explain the experimental
findings shown in Fig.~4.

Although further theoretical investigation is necessary to
enlighten the real nature of the pseudogap state, we believe to
have experimentally proved in a broad doping range ($0.08 \leq x
\leq 0.2$) the existence of two energy scales in LSCO, related to
the separation between a large incoherent pseudogap and a smaller
phase-coherent superconducting gap which follows the
$T_\mathrm{c}$ vs. $x$ behaviour.

The interpretation of these results could play an essential role
in the way to the comprehension of the microscopic mechanism
leading to high-$T_\mathrm{c}$ superconductivity in LSCO.

Many thanks are due to G. Deutscher and A. Perali for useful
discussions. This work has been done under the Advanced Research
Project "PRA-SPIS" of the Istituto Nazionale di Fisica della
Materia (INFM). One of the authors (V.A.S.) also acknowledges the
partial support by the Russian Foundation for Basic Research
(grant No 99-02-17877) and by the Russian Ministry of Science and
Technical Policy within the program ``Actual Problems of Condensed
Matter Physics'' (grant No  96001).

\begin{figure}[t]
\includegraphics[keepaspectratio,width=0.95\columnwidth]{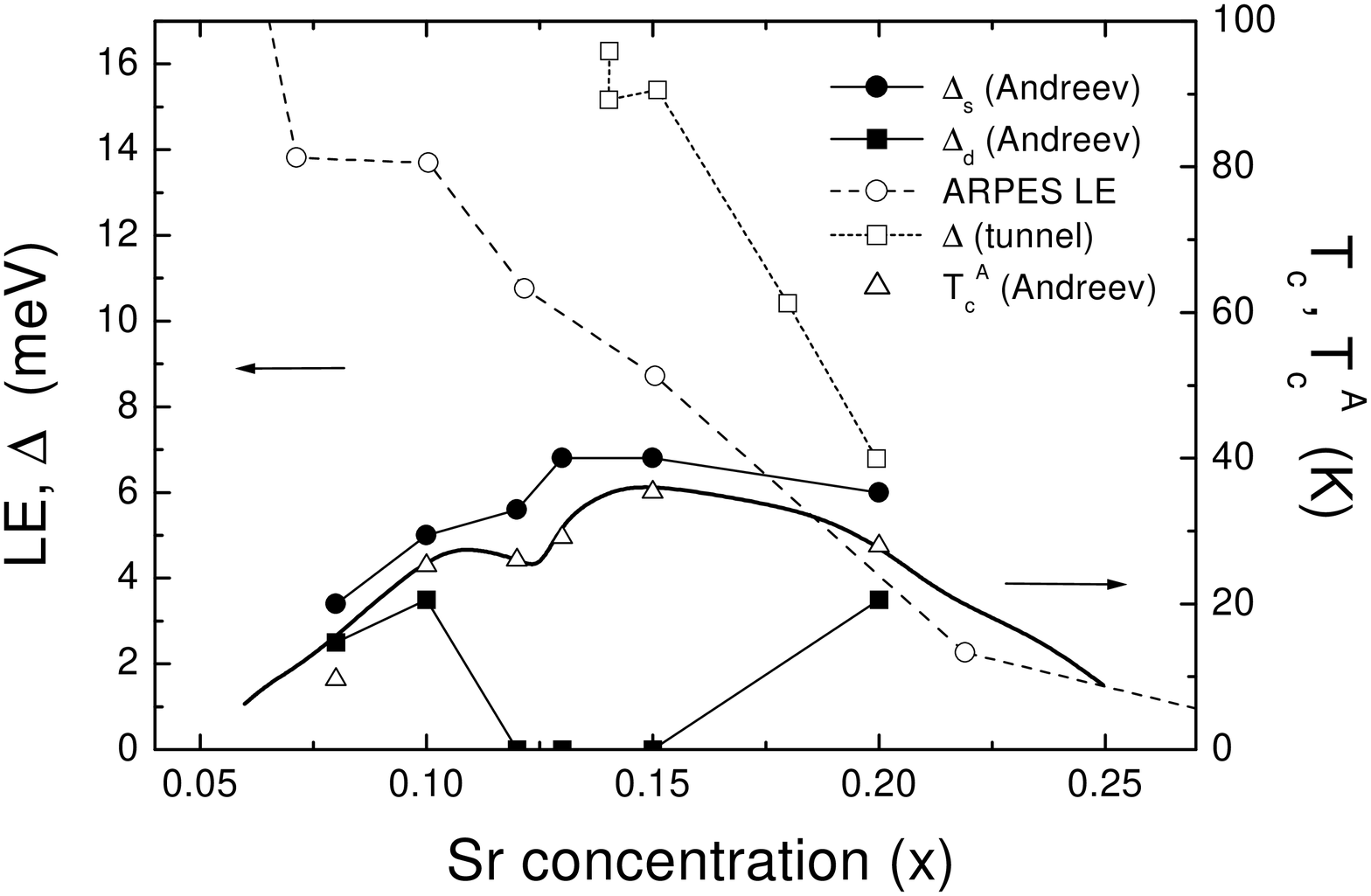}
\vspace{-2mm} \caption{\small{Doping dependence of the ARPES
leading-edge shift (open circles, from Ref.\cite{ref6}), of the
tunneling gap (open squares, from Ref.\cite{ref4}) and of our
point-contact Andreev gap (solid circles for $\Delta_{\mathrm{s}}$
and solid squares for $\Delta_{\mathrm{d}}$) in LSCO. The
temperatures $T_\mathrm{c}^\mathrm{A}$ at which the Andreev
features disappear in our samples are also reported (up triangles)
and compared to the $T_\mathrm{c}$ vs $x$ curve from
Ref.\cite{ref10} (thick line).}}\vspace{-2mm}
\end{figure}

\end{document}